\documentclass[fleqn,usenatbib]{mnras}

\usepackage{newtxtext,newtxmath}

\usepackage[T1]{fontenc}

\DeclareRobustCommand{\VAN}[3]{#2}
\let\VANthebibliography\thebibliography
\def\thebibliography{\DeclareRobustCommand{\VAN}[3]{##3}\VANthebibliography}

\usepackage{graphicx}	
\usepackage{amsmath}	

\usepackage{tikz}
\usetikzlibrary{shapes.geometric, arrows}
\usetikzlibrary{fit}

\tikzstyle{io} = [trapezium, 
trapezium stretches=true, 
trapezium left angle=70, 
trapezium right angle=110, 
minimum width=3cm, 
minimum height=1cm, text centered, 
draw=black, fill=blue!30]

\tikzstyle{process} = [rectangle, 
minimum width=3cm, 
minimum height=1cm, 
text centered, 
text width=3cm, 
draw=black, 
fill=orange!30]

\tikzstyle{decision} = [diamond, 
minimum width=3cm, 
minimum height=1cm, 
text centered, 
draw=black, 
fill=green!30]
\tikzstyle{arrow} = [thick,->,>=stealth]

\makeatletter
\tikzset{
    database/.style={
        path picture={
            \draw (0, 1.5*\database@segmentheight) circle [x radius=\database@radius,y radius=\database@aspectratio*\database@radius];
            \draw (-\database@radius, 0.5*\database@segmentheight) arc [start angle=180,end angle=360,x radius=\database@radius, y radius=\database@aspectratio*\database@radius];
            \draw (-\database@radius,-0.5*\database@segmentheight) arc [start angle=180,end angle=360,x radius=\database@radius, y radius=\database@aspectratio*\database@radius];
            \draw (-\database@radius,1.5*\database@segmentheight) -- ++(0,-3*\database@segmentheight) arc [start angle=180,end angle=360,x radius=\database@radius, y radius=\database@aspectratio*\database@radius] -- ++(0,3*\database@segmentheight);
        },
        minimum width=2*\database@radius + \pgflinewidth,
        minimum height=3*\database@segmentheight + 2*\database@aspectratio*\database@radius + \pgflinewidth,
    },
    database segment height/.store in=\database@segmentheight,
    database radius/.store in=\database@radius,
    database aspect ratio/.store in=\database@aspectratio,
    database segment height=0.1cm,
    database radius=0.25cm,
    database aspect ratio=0.35,
}
\makeatother

\title[24,000~hour GREENBURST search]{Twenty-four thousand hours of GREENBURST observations with the GBT}

\author[J. W. Kania et al.]{
J. W. Kania,$^{1,2,3}$
S. Paine,$^{1,2}$
G. M. Doskoch,$^{1,2}$
S. Tabassum,$^{1,2}$
S. Sirota,$^{1,2}$
M. Flanagan,$^{1,2}$
K. Halley,$^{1,2}$
\newauthor
D. R. Lorimer,$^{1,2}$\thanks{E-mail: duncan.lorimer@mail.wvu.edu (DRL)}
E. Mayfield,$^4$
M. A. McLaughlin,$^{1,2}$
E. Fonseca,$^{1,2}$
D. Agarwal,$^{1,2}$
M. P. Surnis,$^{5}$
F. Crawford,$^{6}$
\newauthor
T. Jespersen,$^{6}$
E. Craver,$^{6}$
M. Golden,$^{6}$
A. Turan,$^{6}$
J. Muyskens,$^{6}$
D. Adair,$^{6}$
Fengqiu Adam Dong,$^{7,8}$
\newauthor
A. P. V. Siemion,$^{9,10,11,3}$
G. Golpayegani,$^{1,2}$
M. B. Mickaliger,$^{3}$
K. M. Rajwade$^{9}$ and
I. H. Stairs$^{12}$
\\
$^{1}$Department of Physics and Astronomy, West Virginia University, Morgantown, WV 26506-6315, USA\\
$^{2}$Center for Gravitational Waves and Cosmology, Chestnut Ridge Building, Morgantown, WV 26505, USA\\
$^{3}$Jodrell Bank Centre for Astrophysics, University of Manchester, Alan Turing Building, Oxford Road M13 9PY, UK\\
$^{4}$Department of Physics and Astronomy, Appalachian State University, Boone, NC 28608-2106, USA\\
$^{5}$Department of Physics, IISER Bhopal, Bhauri Bypass Road, Bhopal, 462066, India\\
$^{6}$Department of Physics and Astronomy, Franklin and Marshall College, P.O. Box 3003, Lancaster, PA 17604, USA \\
$^{7}$National Radio Astronomy Observatory, 520 Edgemont Rd, Charlottesville, VA 22903, USA\\
$^{8}$Green Bank Observatory, 155 Observatory Road, WV 24944, USA\\
$^{9}$Sub-Department of Astrophysics, Department of Physics, University of Oxford, Parks Rd, Oxford OX1 3PU, United Kingdom\\
$^{10}$SETI Institute, 339 N. Bernardo Ave, Mountain View, California 94043, USA\\
$^{11}$Berkeley SETI Research Center, University of California, 339 Campbell Hall, Berkeley, CA 94720, USA\\
$^{12}$Department of Physics and Astronomy, University of British Columbia, 6224 Agricultural Road, Vancouver, BC V6T 1Z1, Canada\\
}

\date{Accepted 2026 April 02. Received 2026 March 19; in original form 2026 January 27}

\pubyear{2026}

\begin{document}
\label{firstpage}
\pagerange{\pageref{firstpage}--\pageref{lastpage}}
\maketitle

\begin{abstract}
In addition to fast radio burst (FRB) searches carried out using dedicated surveys, a number of radio observatories take advantage of commensal opportunities with large facilities in which observations for other projects can be searched for FRBs and other transient sources. We present the results from one such effort, the first 24,186 hours of the GREENBURST search for dispersed radio pulses with the Green Bank Telescope (GBT). To date, GREENBURST has detected a total of 50 pulsars and three FRBs. One of the pulsars, PSR~J0039+5407, has a period of 2.2~s and was previously unknown. Using follow-up observations with the Canadian Hydrogen Intensity Mapping Experiment, we found a timing solution for this pulsar which shows it to have a characteristic age of 2~Myr. Additional GBT observations show the pulsar has a very high nulling fraction ($\sim$70--80\%). All three of the FRBs are repeating sources that were previously known and were being monitored by the GBT as part of other projects. A major challenge for GREENBURST in the discovery of new FRBs is its single beam. This makes it hard to distinguish some of the pulses from sources of radio frequency interference. We highlight this problem with a case study of an FRB-like pulse that initially passed our interference filters. Upon closer inspection, the event appears to be part of a longer-duration narrow-band source of unknown origin. Further observations and monitoring are required to determine whether it is terrestrial or
celestial.
\end{abstract}




\section{Introduction}

As a population of cosmological sources of unknown origin, fast radio bursts (FRBs) are insightful probes of fundamental physics and the large-scale structure of the Universe \citep[for a recent review, see][]{2023RvMP...95c5005Z}. Among the many \cite[for a living compilation, see][]{2019PhR...821....1P} possible sources {suggested} to explain FRBs so far are extragalactic magnetars and compact object mergers. The discovery of repeating FRBs \citep{2016Natur.531..202S,2019Natur.566..235C} suggests that at least some FRBs are produced in non-catastrophic events. The discovery of FRB-like emission from the Galactic magnetar SGR~1953+2154 \citep{2020Natur.587...59B,2020Natur.587...54C} provides strong evidence in favour of a magnetar origin for some fraction of the FRB population. The discoveries of FRBs in galaxies at increasingly larger redshifts \citep{2023Sci...382..294R,2025arXiv250801648C} highlight the need to characterize this population using high-sensitivity instruments.

Motivated by the need to find more FRBs, as part of a collaboration with groups engaged in the search for extraterrestrial intelligence \citep{2017ApJS..228...21C} using the SERENDIP VI backend architecture \citep{serendip_6}, we began building and commissioning real-time commensal FRB detectors on the Arecibo \citep[ALFABURST;][]{2018MNRAS.474.3847F} and Green Bank \citep[GREENBURST;][]{2019PASA...36...32S} telescopes. Since the demise of the Arecibo telescope, our focus has been on GREENBURST, with commissioning and initial observations being reported by \citet{2020MNRAS.497..352A}. A key component of that work was our development \citep{2020MNRAS.497.1661A} of the Fast Extragalactic Transient Candidate Hunter (FETCH), a deep-learning model that autonomously identifies FRBs. Some of the development of FETCH, particularly in the form of early training
data, was provided by GREENBURST commissioning observations.

The goal of GREENBURST is to detect dispersed radio pulses of astrophysical origin by piggybacking on Green Bank Telescope (GBT) observations. Since 2019, GREENBURST has collected over 20,000 hours of GBT data as part of its routine operations, increasing the scientific productivity of the telescope and providing a basis for PhD projects \citep{agarwal2020searches,golpayegani2020searches,kania2023applications}. We describe these observations in this work which has so far led to the successful detections of FRBs and pulsars as well as the discovery of one Galactic pulsar through the detection of individual pulses. 

The plan for the rest of this paper is as follows. In \S 2, we briefly describe the observational framework of GREENBURST as well as infrastructure developments made since the description provided by  \citet{2020MNRAS.497..352A}. In \S 3, we detail our main results to date. The implications of these results are discussed further in \S 4 before we conclude with some projections for the future in \S 5.

\begin{figure*}
\begin{center}

\begin{tikzpicture}[scale=1.2, transform shape, node distance=2cm]

\node (in1) [io] {UDP Packets};
\node (pro1) [process, right of=in1,  xshift=1.5cm] {Double Buffer};
\node (pro2) [process, right of=pro1, xshift=1.5cm] {Filterbank};
\node (dec1) [decision, below of=pro2] {Data Valid?};
\node (in2) [io, below of=in1,] {Telescope Metadata};
\node (db1) [database, right of=in2,database radius=0.6cm, xshift=1.5cm, database segment height=0.3cm, label=below:InfluxDB] {};
\coordinate [below of=db1, yshift=1cm] (coord1);
\node (db2) [database, below of=db1,database radius=0.6cm, database segment height=0.3cm, label=below:Elasticsearch] {};
\node (pro3) [process, right of=dec1, xshift=1.6cm, fill=yellow!60] {jess\_gauss.py};
\node (pro4) [process, below of=pro3, fill=yellow!60] {Heimdall};
\node (pro5) [process, below of=pro4] {Filter Candidates};
\node (pro6) [process, below of=pro5, fill=yellow!60] {Candidate Cutouts};
\node (dec2) [decision, left of=pro6, fill=yellow!60, xshift=-1.6cm] {Fetch};
\node (dec3) [decision, left of=dec2, xshift=-1.6cm] {Manual Inspect};
\node (pro7) [process, left of=pro5, xshift=-1.6cm] {Delete};
\node (pro8) [process, left of=dec3, xshift=-1.6cm] {Followup};


\draw [arrow] (in1) -- (pro1);
\draw [arrow] (pro1) -- (pro2);
\draw [arrow] (pro2) -- (dec1);
\draw [arrow] (in2) -- (db1);
\draw [arrow] (db1) -- (dec1);
\draw [arrow] (coord1) -- (db2);
\draw [arrow] (dec1) -- node[pos=0.5,above]{Yes}(pro3);
\draw [arrow] (pro3) -- (pro4);
\draw [arrow] (pro4) -- (pro5);
\draw [arrow] (pro5) -- (pro6);
\draw [arrow] (pro6) -- (dec2);
\draw [arrow] (dec2) -- node[pos=0.5,above]{Yes}(dec3);
\draw [arrow] (dec2) -- node[pos=0.5,left]{No}(pro7);
\draw [arrow] (dec1) -- node[pos=0.5,left]{No}(pro7);
\draw [arrow] (dec1) -- node[pos=0.5,left]{No}(pro7);
\draw [arrow] (dec3) -- node[pos=0.5,above]{Yes}(pro8);


\end{tikzpicture}
   \end{center}
    \caption{Schematic showing the GREENBURST   data acquisition and analysis pipeline. The blue input squares are data provided by the Serendip VI \citep{serendip_6} backend. Data are sent to GREENBURST via UDP packets which are buffered and written in {\tt filterbank} format \citep{2011ascl.soft07016L}. The double buffer ensures continuous data coverage over discrete filterbanks. The Elasticsearch database keeps track of the telescope pointing at 8~min cadence over the lifetime of the experiment. If the data are valid (suitable receiver position, not in maintenance, etc.) they are cleaned with \textsc{jess\_gauss.py}’s Jarque-Bera filter. These cleaned filterbank data are subsequently searched with {\tt heimdall} \citep{Barsdell}. Candidate pulses are cut out and classified as astrophysical or RFI by FETCH \citep{2020MNRAS.497.1661A}. Candidates classified as astrophysical  are reviewed by humans on a daily basis via a dedicated {\tt Slack} channel. An InfluxDB2 keeps track of the telescope pointing at a cadence of 1~s. Yellow squares show steps that leverage GPU acceleration.}
    \label{fig:schematic}
\end{figure*}
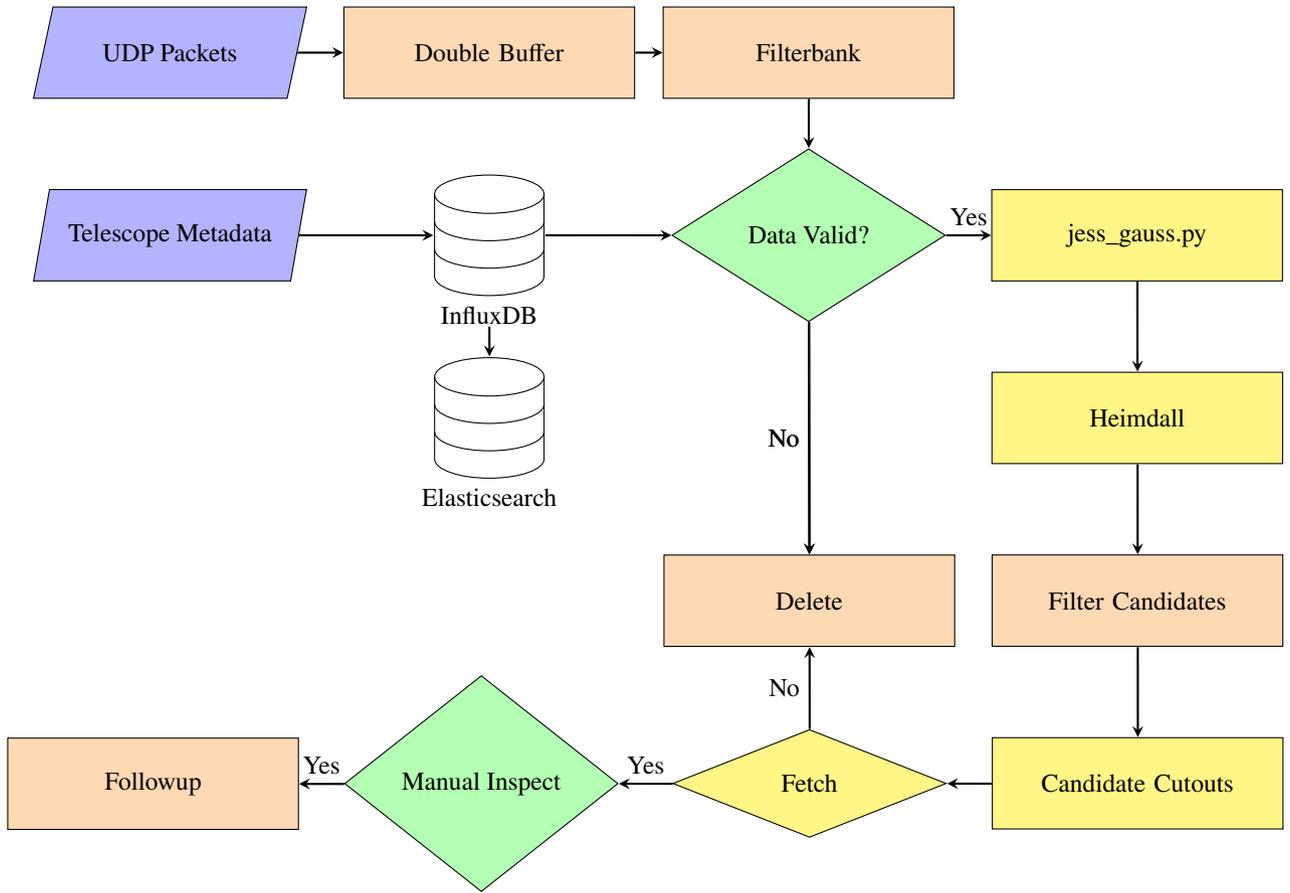

\section{Observational overview and updates}\label{sec:obsover}

As described in detail in \citet{2019PASA...36...32S}, GREENBURST and SERENDIP VI \citep{serendip_6} share a copy of the signal from the L-band (1.4~GHz) feed from the GBT. A schematic of the resulting data flow and analysis pipeline is shown in Fig.~\ref{fig:schematic}. The GREENBURST signal is 4096 channels covering 960~MHz, centred at 1440~MHz sampled at 256~$\mu$s with 8-bit precision. These data are searched for the presence of dispersed pulses of astrophysical origin with a pipeline optimized for the detection of pulses with dispersion measures  (DMs) from 10 to 10,000~pc~cm$^{-3}$ and widths from 256~$\mu$s to 32.8~ms. Further details of the data analysis pipeline can be found in
\citet{2020MNRAS.497..352A}. The major change to the pipeline since this earlier account has been the implementation \citep{kania2023applications} of more sophisticated techniques to remove unwanted radio frequency interference (RFI) signals, as described below.

Radiometer noise and signals from astronomical sources are expected to be normally distributed \citep{Gary_2007, Nita_2010}. Communication systems are non-thermal, and therefore are often highly non-Gaussian, and we expect communication signals, one of the main sources of RFI,  to have higher statistical moments \citep{Rafiei-Ravandi_2023}. We therefore use normality tests to discriminate between astrophysical and anthropogenic signals. Spectral kurtosis (the fourth normalized moment) has been shown to be an excellent discriminator for RFI, and subject of much development by the radio astronomy community \citep{Gary_2007, Gary_2010, Nita_2010, Taylor_2019, Nita_2019, Mirhosseini_2020, Smith_2022}. Kurtosis measures the tails of a distribution, making it insensitive to signals with a 50\% duty cycle. This insensitivity has traditionally been mitigated by using multi-scale kurtosis tests \citep{Gary_2010, Smith_2022}. 

As will be described in detail in an upcoming paper (Kania et al., in preparation), we investigated alternative normality tests which involve both kurtosis and skew (third normalized moment). For this paper, we look for outliers in kurtosis and skew individually, then combine the resulting masks. We look for outliers in the Jarque–Bera  statistic \citep{Jarque_1980, Bera_1981, Jarque_1987} or D'Agostino's $K^{2}$ statistic \citep{DAgostino_1973, DAgostino_1990}. Outliers are identified by calculating the inter-quartile range (IQR) and flagging samples that are four standard deviations away from the median. IQR is a robust measure of scale \citep{Rousseeuw_1992,2022MNRAS.510.1393M}, making it resilient to RFI-contaminated sections of spectra. We consider blocks of 64 and 4096 samples (corresponding to approximately 16 ms and 1 s, respectively). Additionally, following the normality-based excision, we flag sections on a 16384 sample (4.2~s) cadence with unusually high median versus mean absolute difference. While we expect RFI to be more prominent at higher moments, some RFI seems to be particularity strong driving the signal chain into a non-linear state. These non-linear sections do not show up more significantly at
higher statistical moments, so we perform this first-moment normality test. Once we have excised sections of dynamic spectra based on their normality, we use the high-pass filter first described in \citet{eatough-2009} and further refined by \citet{lazarus}. \citet{eatough-2009} describes a technique to remove broadband signals by taking the mean across all frequency channels, creating a zero-DM time series then subtracting this time series from each frequency channel. 
This works well when there is uniform frequency response. Often, telescopes do not have uniform sensitivity across their bandpass, due to receiver roll-off and band-stop filter. To overcome this changing sensitivity, \citet{lazarus} introduce a weighted 
zero-DM subtraction technique that uses the bandpass shape to account for this changing receiver sensitivity by using a weighted bandpass. As described in \citep{millisecond_filters}, we use this technique to remove broadband RFI in addition to the noise diode. 

\section{Results}

\begin{figure*}
    \centering
    \includegraphics[width=0.9\textwidth]{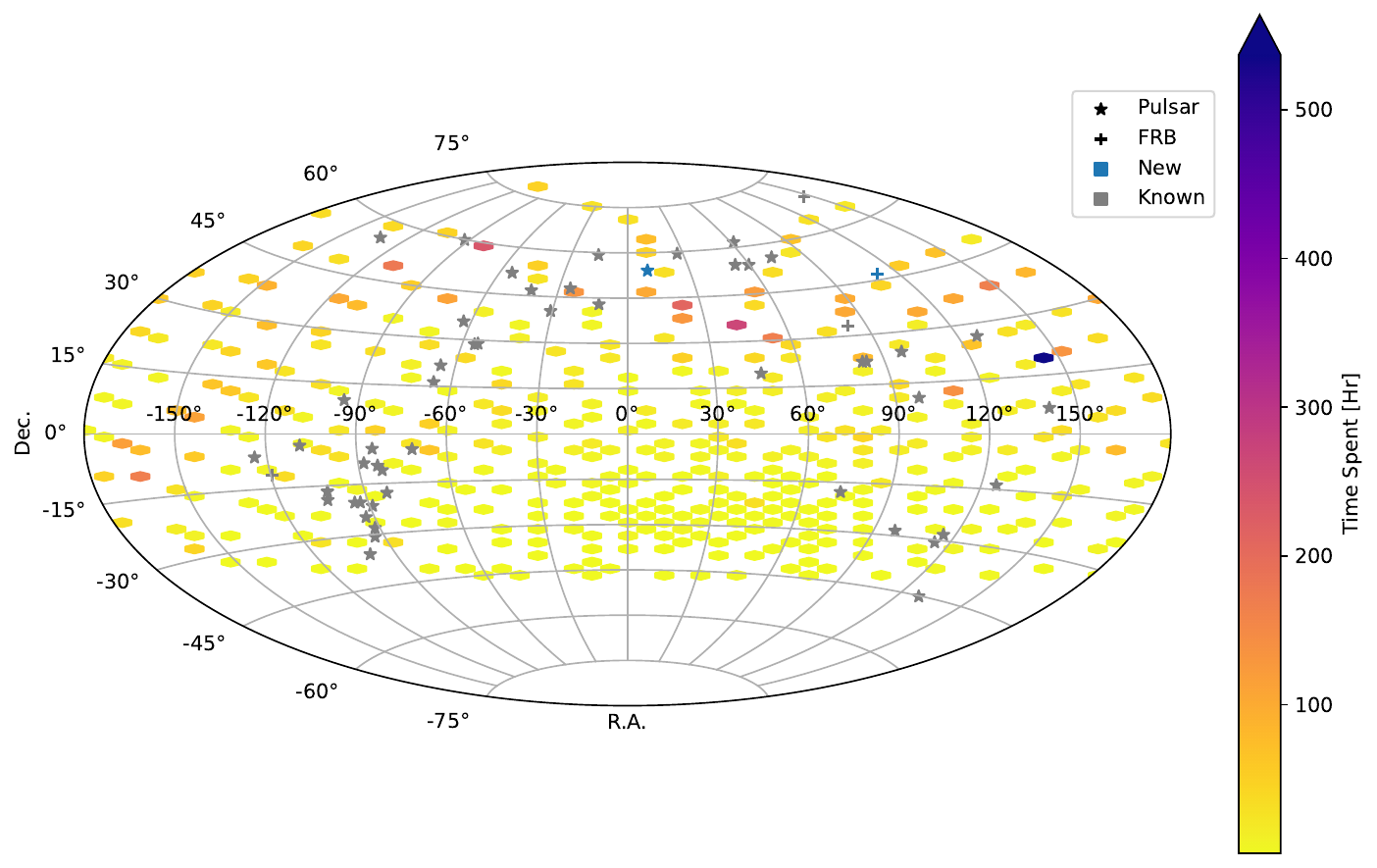}
    \caption{Sky coverage for the observation period reported in this paper. The positions are quantized into 6~deg$^2$ hexagons and colour coded into hours spent per region. Stars indicate pulsars, crosses indicate FRBs. Grey points are known objects. The blue star is represents PSR~J0039+5407. The blue cross is GBP~220718.}
    \label{fig:hours_spent}
\end{figure*}

\begin{figure}
    \includegraphics[width=0.49\textwidth]{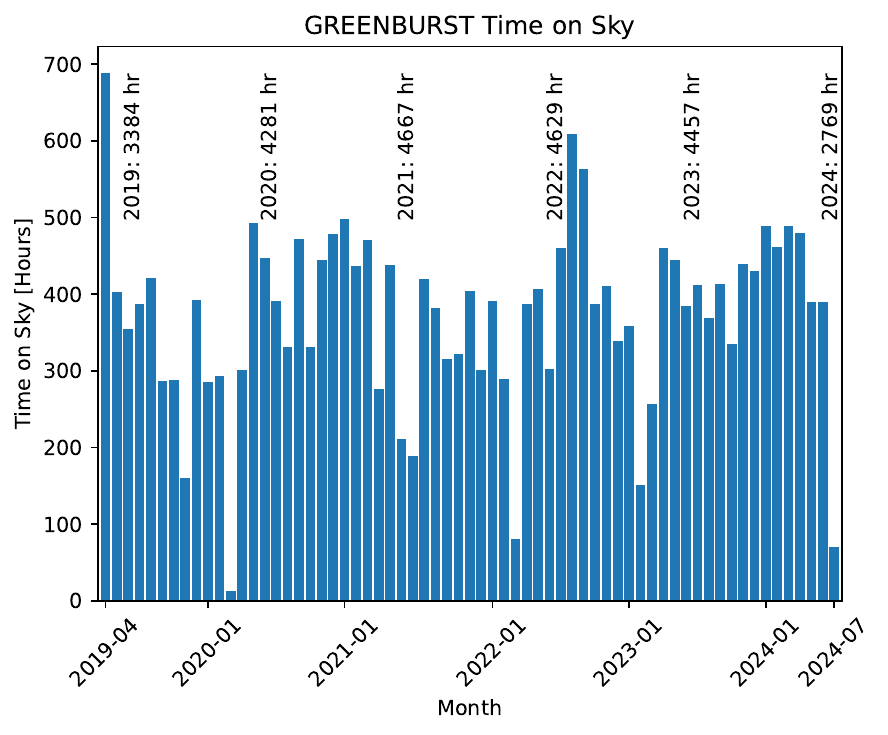}
    \caption{Timeline of observation showing totals for each year. The 2024 total is up to the end of July. Overall, the L-Band receiver was in focus 72\% of the time. Other receivers being used by the primary observer were: X-Band 12\%, C-Band 7\%, Ka-Band 5\% and Mustang 3\%.}
    \label{fig:time_spent_hist}
\end{figure}

The results from this work cover GBT observations between March 14, 2019 and July 31, 2024 during which time a total of 24,186 hours of data were collected.
Fig.~\ref{fig:hours_spent} shows an equatorial projection of the sky coverage to date. Fig.~\ref{fig:time_spent_hist} shows the time spent on sky 
as a function of epoch. In the subsections below, we detail the main outcomes.

\begin{figure*}
    \centering
     \includegraphics[width=0.9\textwidth]{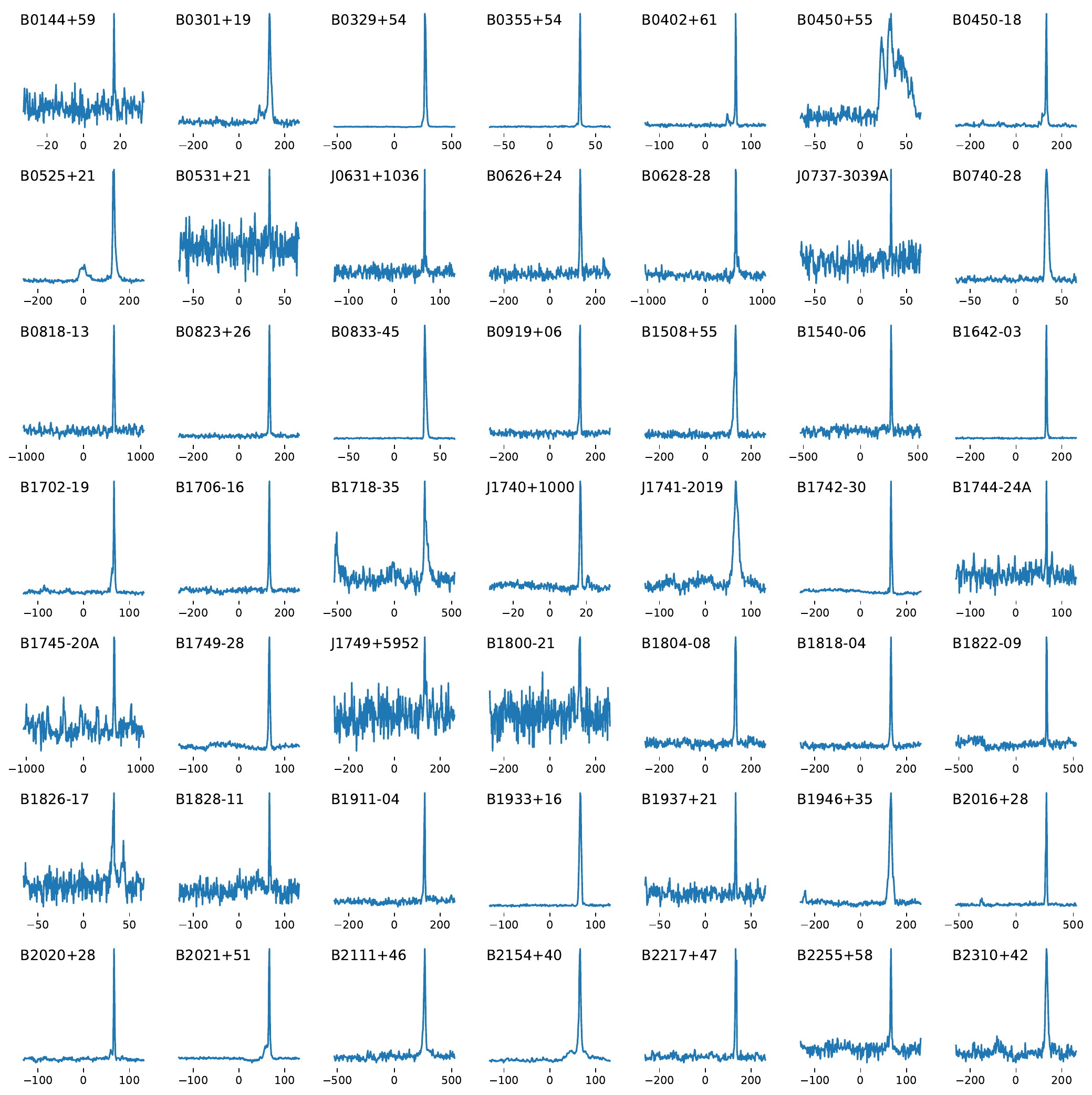}
     \caption{Sample individual pulses detected from the 49 pulsars in Table 1. The numerical labels under each pulse denote time in ms with respect to the centre of the time series. For presentation purposes, each pulse has been aligned so that its peak value is located 75\% along the horizontal axis.}
     \label{fig:knownPSRs}
\end{figure*}

\begin{figure*}
     \includegraphics[width=0.24\textwidth]{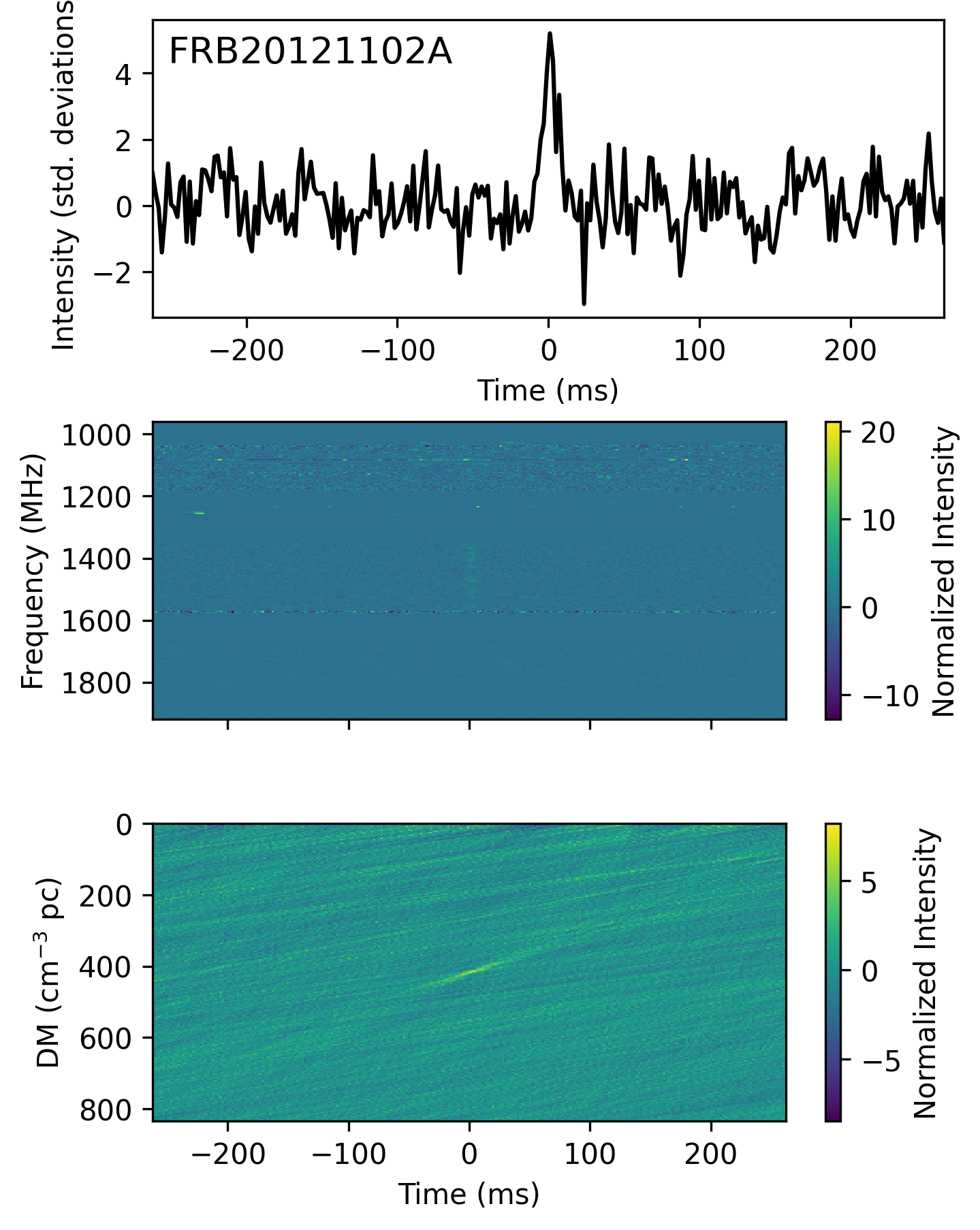}
     \includegraphics[width=0.24\textwidth]{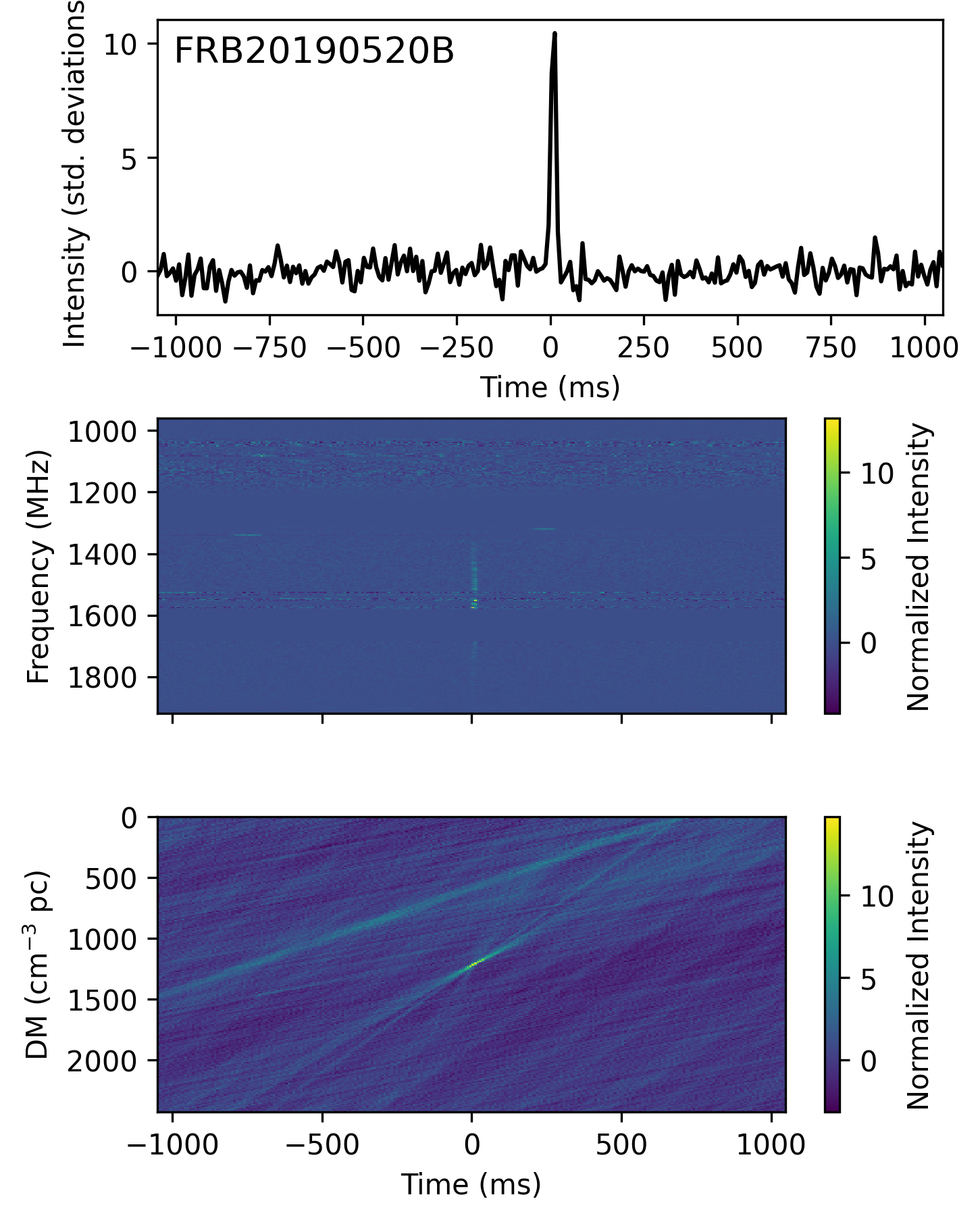}
     \includegraphics[width=0.24\textwidth]{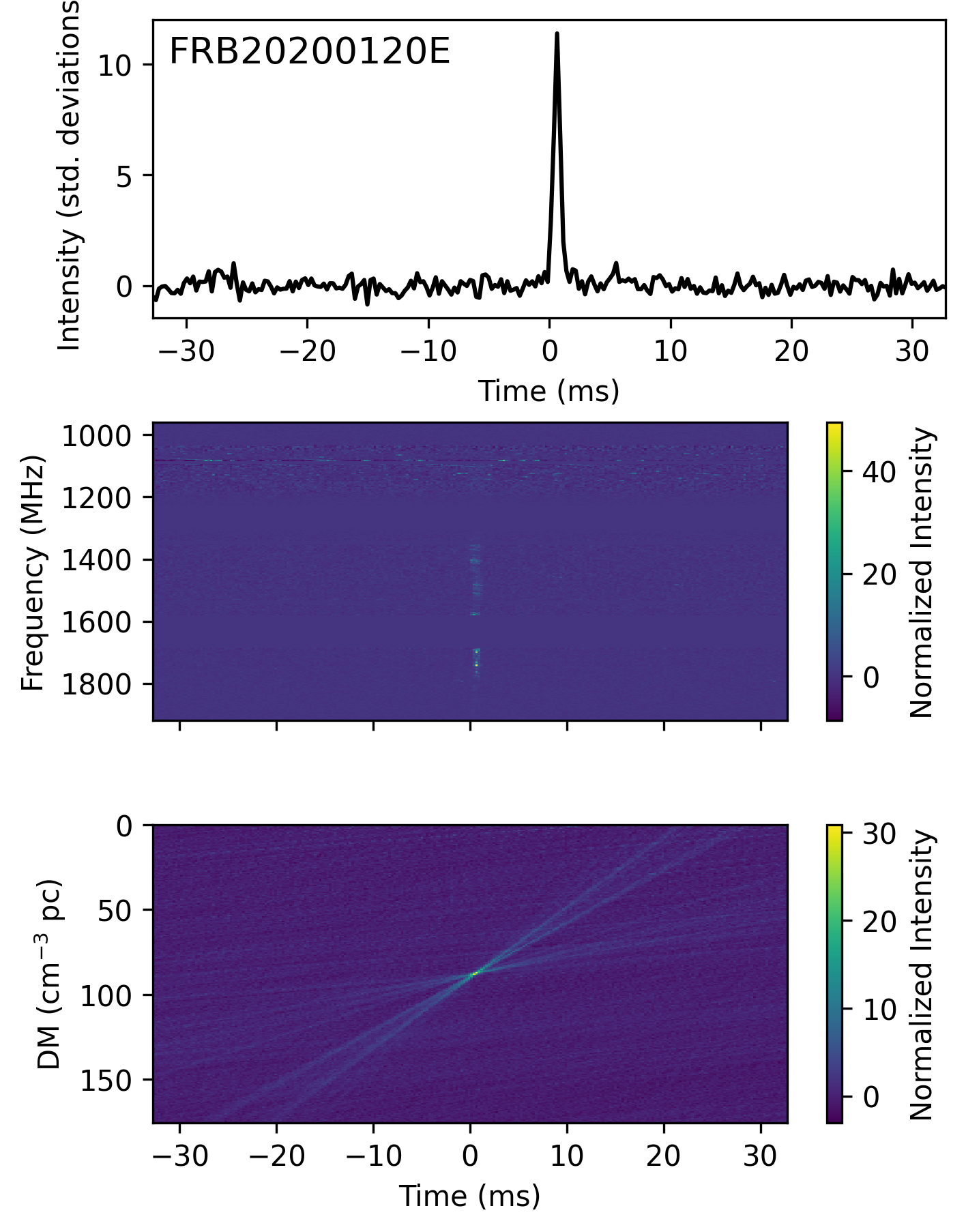}
     \includegraphics[width=0.24\textwidth]{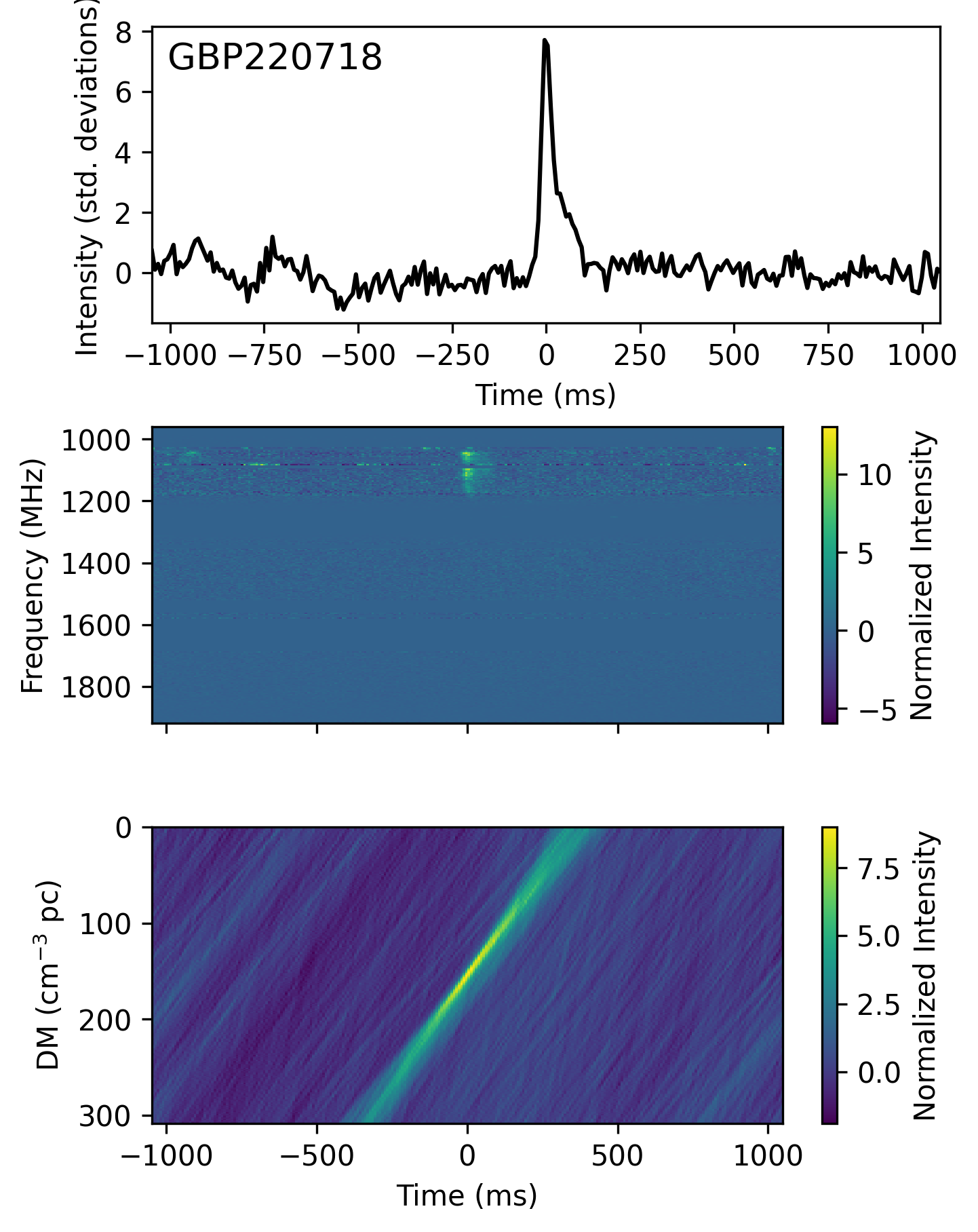}
     \caption{Example detections of individual pulses from the three previously known FRBs
     (FRB~20121102A, FRB~20190520B, and FRB~20200120E) observed during follow-up observations of these sources as well as the discovery pulse for GBP~220718. In each case the top panel shows the dedispersed pulse, the middle panel shows the `waterfall' plot of frequency vs time, while the bottom panel shows the trial DM as a function of time.  In each of the data visualizations, the median has been subtracted from the data and the resulting arrays have been divided by their respective standard deviation.}
     \label{fig:knownFRBs}
\end{figure*}

\subsection{Detections of previously known sources}

Our observations to date have resulted in the detection of 16,556 pulses from 49 previously known pulsars. These detections are summarized in Table \ref{tab:knownPSRs} and the brightest pulse from each source is shown in Fig.~\ref{fig:knownPSRs}. While further analyses of these pulses will
be provided in a subsequent publication (Tabassum et al., in preparation), some notable detections among this sample are: bright pulses from the 11~ms
pulsar B1744--24A in Terzan~5 \citep[which are characterized in earlier GBT observations by][and thought to be due to plasma lensing in this eclipsing
binary system]{2019ApJ...877..125B}, giant pulses from the first millisecond pulsar B1937+21 \citep{1996ApJ...457L..81C}, and individual pulses
from the 22.7~ms pulsar J0737--3039A in the double pulsar system \citep{2003Natur.426..531B,2004Sci...303.1153L}.

\begin{table}
\label{tab:knownPSRs}
\centering
\caption{The 49 previously known pulsars detected during GREENBURST operations. For each pulsar, we list the spin period ($P$), dispersion measure (DM), 
1400~MHz flux density ($S_{1400}$), number of epochs ($N_{\rm epochs}$) on which the source was detected, and the total number of pulses detected  ($N_{\rm pulses}$).}
\begin{tabular}{@{}lrrrrr}
\hline
PSR & \multicolumn{1}{c}{$P$} & \multicolumn{1}{c}{DM} & \multicolumn{1}{c}{$S_{1400}$} & \multicolumn{1}{c}{$N_{\rm epochs}$} & \multicolumn{1}{c}{$N_{\rm pulses}$}  \\
    & \multicolumn{1}{c}{(ms)}& \multicolumn{1}{c}{(cm$^{-3}$~pc)} & \multicolumn{1}{c}{(mJy)}  &  &    \\
\hline
B0144+59 & 196.32 & 40.1 & 2.1 & 1 & 4  \\
B0301+19 & 1387.58 & 15.7 & 15.0 & 3 & 4  \\
B0329+54 & 714.52 & 26.8 & 203.0 & 31 & 837  \\
B0355+54 & 156.38 & 57.1 & 23.0 & 96 & 4408  \\
B0402+61 & 594.58 & 65.4 & 2.8 & 1 & 17  \\
B0450$-$18 & 548.94 & 39.9 & 17.0 & 2 & 557  \\
B0450+55 & 340.73 & 14.6 & 13.0 & 1 & 1  \\
B0525+21 & 3745.54 & 50.9 & 8.9 & 3 & 13  \\
B0531+21 & 33.39 & 56.8 & 14.0 & 1 & 1  \\
B0626+24 & 476.62 & 84.2 & 3.2 & 5 & 107  \\
B0628$-$28 & 1244.42 & 34.4 & 32.0 & 1 & 91  \\
J0631+1036 & 287.82 & 125.3 & 1.1 & 1 & 10  \\
J0737$-$3039A & 22.69 & 48.9 & 1.3 & 2 & 3  \\
B0740$-$28 & 166.76 & 73.7 & 26.0 & 14 & 314  \\
B0818$-$13 & 1238.13 & 40.9 & 6.0 & 3 & 169  \\
B0823+26 & 530.66 & 19.5 & 10.0 & 1 & 164  \\
B0833$-$45 & 89.33 & 67.8 & 1050.0 & 1 & 146  \\
B0919+06 & 430.63 & 27.3 & 10.0 & 1 & 32  \\
B1508+55 & 739.68 & 19.6 & 8.0 &5& 44  \\
B1540$-$06 & 709.06 & 18.3 & 2.0 &1& 110  \\
B1642$-$03 & 387.69 & 35.8 & 25.8 &1&  338 \\
B1702$-$19 & 298.99 & 22.9 & 5.7 & 52 & 4189  \\
B1706$-$16 & 653.05 & 24.9 & 15.0 & 1 & 162  \\
B1718$-$35 & 280.42 & 496.8 & 16.8 & 4 & 540  \\
J1740+1000 & 154.10 & 23.9 & 2.7 & 1 & 2  \\
J1741$-$2019 & 3904.51 & 74.9 & 0.0 &2& 2  \\
B1742$-$30 & 367.43 & 88.4 & 21.0 & 3 & 66  \\
B1745$-$20A & 288.60 & 219.4 & 0.4 & 1 & 2  \\
B1744$-$24A & 11.56 & 242.2 & 0.6 & 3 & 13  \\
J1749+5952 & 436.04 & 45.1 & 0.0 & 1 & 1  \\
B1749$-$28 & 562.56 & 50.4 & 48.0 & 3 & 3  \\
B1800$-$21 & 133.67 & 234.0 & 9.6 & 1 & 8  \\
B1804$-$08 & 163.73 & 112.4 & 18.0 &2 & 108  \\
B1818$-$04 & 598.08 & 84.4 & 10.1 &1& 18  \\
B1822$-$09 & 769.02 & 19.4 & 10.0 &3 & 6  \\
B1826$-$17 & 307.13 & 216.8 & 11.0 & 1 & 3  \\
B1828$-$11 & 405.07 & 159.7 & 1.5 &15 & 1216  \\
B1911$-$04 & 825.94 & 89.4 & 6.8 &1& 53  \\
B1933+16 & 358.74 & 158.5 & 58.0 & 15 &  1001 \\
B1937+21 & 1.56 & 71.0 & 13.9 & 21 & 33  \\
B1946+35 & 717.31 & 129.4 & 8.3 & 4 & 98  \\
B2016+28 & 557.95 & 14.2 & 30.0 & 1 & 5  \\
B2020+28 & 343.40 & 24.6 & 38.0 & 6 &  640 \\
B2021+51 & 529.20 & 22.6 & 27.0 &36&  828 \\
B2111+46 & 1014.69 & 141.2 & 19.0 & 1 &  4 \\
B2154+40 & 1525.27 & 71.1 & 17.0 &1&  21 \\
B2217+47 & 538.47 & 43.4 & 3.0 &2& 67  \\
B2255+58 & 368.25 & 151.1 & 9.2 & 1 &39\\
B2310+42 & 349.43 & 17.3 & 15.0 &4  &61   \\
\hline
\end{tabular}
\end{table}

\subsection{Previously known FRBs}

During the observing period reported here, three repeating FRBs were observed by the GBT for which we have detections: FRB~20190520B \citep{2022Natur.606..873N} was observed as part of project GBT20B\_401, FRB~20200120E \citep{2021ApJ...910L..18B} was observed as part of project GBT22A\_502, and FRB~20121102A \citep{2016Natur.531..202S} was observed as part of project GBT21A\_417. Sample pulses from these sources are shown in Fig.~\ref{fig:knownFRBs}. As these data were collected for other projects with specific scientific goals, we do not use them for any purposes beyond their value as a test of the ability of our data analysis pipeline to make serendipitous detections of FRB pulses.

\subsection{PSR~J0039+5407}

On 9 June 2022 at 10:14:34 UTC, we detected a pulse with a signal-to-noise ratio (S/N) of 14 and a DM of 74.2~cm$^{-3}$~pc. Upon RFI cleaning using the composite filter in \textsc{jess} \citep{millisecond_filters}, eight additional lower-significance pulses were detected.  The initial pulse was found again at S/N = 29, corresponding to a peak flux density of about 70~mJy, as well as three additional pulses in the 10--30~mJy range. Manual inspection of candidates with DMs around 70~cm$^{-3}$~pc revealed another five pulses during the time the pulsar was drifting through the telescope's field of view. From these nine pulses, we were able to determine the most probable position of the pulsar and identify its period. Using a kernel density estimation \citep[KDE;][]{Rosenblatt_1956, Parzen_1962} approach, the most probable position of the pulsar was found. As input to the KDE process, we used a Gaussian smoothing kernel that has a width equal to the telescope beam, \citep[9.2~arcmin;][]{2019PASA...36...32S}. We then used the S/N of the detected pulses as relative weights for the kernels using scikit-learn's KDE implementation \citep{scikit-learn}. This yielded an initial estimate of the right ascension (J2000) consistent with a point source drifting through the beam. The declination uncertainty is simply the width of the telescope beam. A standard Fourier transform-based search for periodicities in these data using PRESTO \citep{Ransom-2011} did not reveal any significant candidates. A fast folding algorithm search using {\tt riptide} \citep{Morello_2020}, however, revealed the presence of a 2.2~s periodicity consistent with the presence of a radio pulsar with that period. These initial parameters were subsequently refined during the follow-up timing analysis for the pulsar (henceforth referred to as PSR~J0039+5407) described in Section \ref{sec:0039followup}.

\subsection{GBP~20220718}\label{sec:gbpdisc}

On 18 July 2022 at 16:07:17 UTC, we detected a pulse of width 16~ms with a DM of 145.5~cm$^{-3}$~pc and S/N of 10. From the position of the telescope at the time of detection, we estimate the right ascension to be $07^{\text{h}}\,25^{\text{m}}\,13^{\text{s}} \pm 18^{\text{s}}$ and the declination to be $+46^{\circ}\,31.8^{\prime}\pm 4.5^{\prime}$. As shown in the discovery plot
in Fig.~\ref{fig:knownFRBs}, the signal was limited to the lower quarter of the bandpass. In our discussion below, due to additional emission that we subsequently found in the original observation, we do not feel confident in currently concluding that it is celestial in origin and henceforth refer to it as GBP~220718, where GBP stands for GREENBURST pulse. 

\section{Discussion}

In the following sections, we discuss the findings from follow-up observations of both PSR~J0039+5407 and GBP~220718.

\subsection{PSR~J0039+5407}\label{sec:0039followup}

\begin{figure}
\centering
    \includegraphics[width=0.425\textwidth]{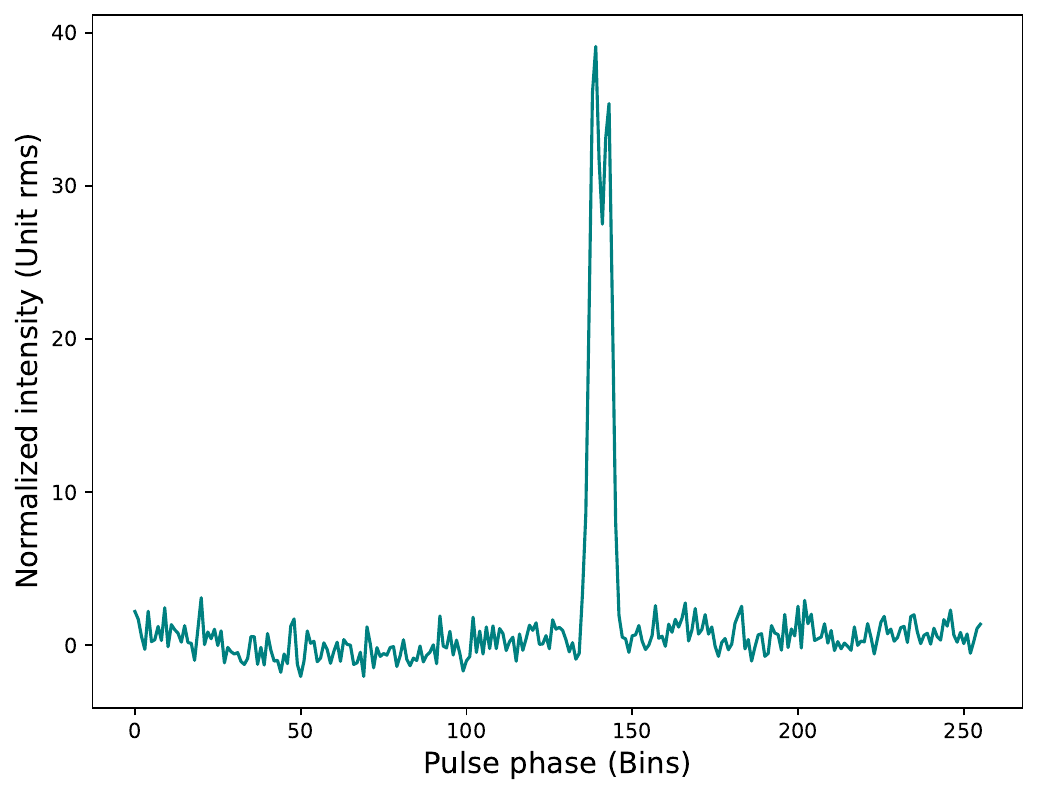}
    \caption{Composite profile of J0039+5407 from the phase-coherent addition of the best CHIME/Pulsar detections. The total integration time is 4263~s.}
    \label{fig:J0039+5407-folded-profile}
\end{figure}

Starting from the initial measurements of the period, DM and position obtained from the discovery observations presented above we carried out regular observations of this pulsar using CHIME/Pulsar \citep{2021ApJS..255....5C}. Using a preliminary ephemeris for this pulsar made from initial CHIME/FRB \citep{2018ApJ...863...48C} observations in search mode in July 2022, a further 134 timing mode observations of 1430~s were taken by CHIME between August 16, 2022 and March 6, 2024. Carrying out a standard timing analysis using the {\tt tempo2} software package \citep{2006MNRAS.369..655H} yields a phase-connected solution spanning 446 days with parameters summarized in Table~\ref{tab:timing}. The integrated profile obtained by summing all of the CHIME/PSR observations is shown in Fig.~\ref{fig:J0039+5407-folded-profile}. The DM fit to
the TOAs was carried out by generating eight 50~MHz sub-bands from one of the strongest CHIME/Pulsar detections and keeping the other parameters in the solution fixed.

\subsubsection{Nulling analysis} \label{subsubsec:j0039+5407-nulling}

Some pulsars exhibit nulling, a pattern of intermittent intervals during which the pulsar enters a state of minimal to no emission \citep{Backer1970}. Nulling can occur on a wide variety of timescales \citep{Wang+2007}, depending on the pulsar. Previous studies have reported correlations between the fraction of null pulses (the null fraction, NF) and pulse period $P$ \citep{Biggs1992} and characteristic age $\tau$ \citep{Ritchings1976,Wang+2007}, with the latter relationship being even stronger. With $P\simeq2.24$~s and $\tau\simeq2.3$~Myr, it would not be surprising if PSR~J0039+5407 exhibited nulling.

\begin{figure}
    \centering
    \includegraphics[width=0.45\textwidth]{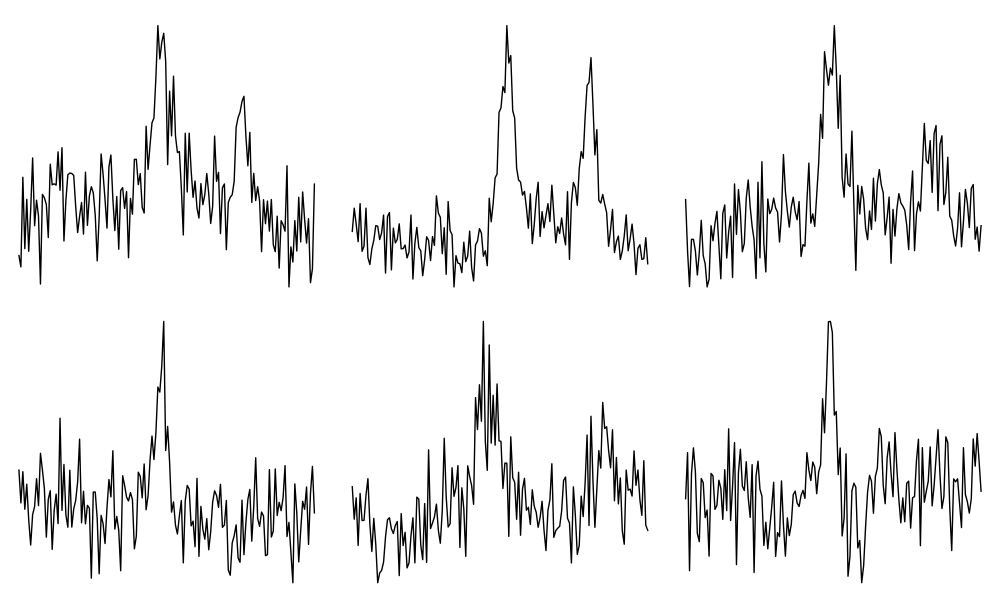}
    \caption{Six single pulses from a GBT 820 MHz observation of J0039+5407 exhibiting several types of morphology. Some pulses show only one peak while others show both. The trailing peak of multi-component pulses is weaker. The horizontal extent of each of the pulses shown is 100~ms and the time series has been downsampled by a factor of 16.}
    \label{fig:J0039+5407-single-pulses}
\end{figure}

We conducted an analysis to find NF for PSR~J0039+5407 using a GBT observation of the pulsar carried out at a central frequency of 820 MHz using the VEGAS spectrometer \citep{2015ursi.confE...4P} in search mode using 2048 channels over a total bandwidth of 200~MHz sampled every 80~$\mu$s. Using the timing solution, we folded the data into single pulses. Fig.~\ref{fig:J0039+5407-single-pulses} shows a selection of the strongest single pulses. An examination of the folded profile showed that the majority of emission is concentrated within a span of $\sim0.1$~s, which we designate the on-pulse region. Following the method of \citet{Brook2019}, after subtracting a baseline from each pulse, we summed the time series over the on-pulse region, yielding some fluence $F$. For the purposes of the analysis below, the fluence scale does not require calibration. However, from radiometer noise considerations \citep[see, e.g.,][]{2004hpa..book.....L}, we find typical peak flux densities on the order of 1~Jy in this observing band
(720--920~MHz).

\begin{figure}
\centering
    \includegraphics[width=0.45\textwidth]{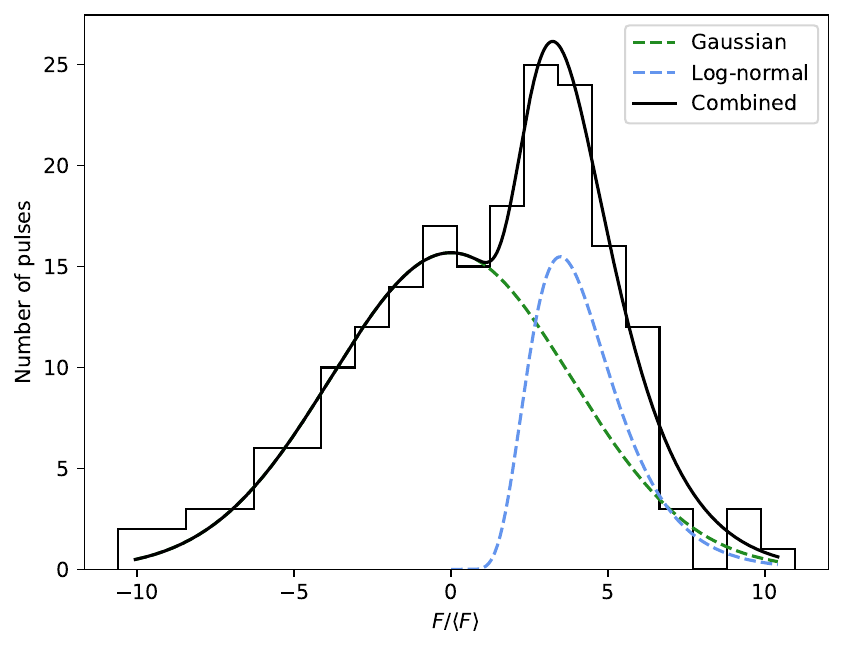}
    \caption{Histograms of normalized single-pulse fluences of J0039+5407, along with the two-component fits. The Gaussian is centered at zero and represents spin periods dominated by noise (the nulls) while the log-normal distribution represents spin periods with stronger emission.}
    \label{fig:J0039+5407-nulling-histogram-fits}
\end{figure}

Assuming Gaussian noise, the pulse periods from nulling should yield fluences following a Gaussian distribution centered at $F=0$. The pulse periods with strong single-pulse emission, on the other hand, should follow a log-normal distribution \citep{BurkeSpolaor2012}.\footnote{While some pulsars follow a slightly different distribution \citep{Mickaliger2018}, a log-normal approximation should be sufficient for our purposes.} We can compute the nulling fraction by fitting a sum of a Gaussian (centered at $F/\langle F\rangle=0$) and log-normal distribution to the empirical flux density histogram. If the two component distributions have fitted amplitudes $A_n$ and $A_e$, respectively, then 
\begin{equation}
{\rm NF}=\frac{A_n}{A_n+A_e},
\end{equation}
where we have ensured that each distribution, before fitting, was normalized to have unit area; in other words, $A_n$ and $A_e$ are also the areas under the distributions. We performed the fits using the \texttt{optimize.curve\_fit} routine from the \texttt{scipy}\footnote{\url{https://scipy.org/}}  package \citep{SciPy}. The fitted histograms are shown in Fig.~\ref{fig:J0039+5407-nulling-histogram-fits}, and yielded ${\rm NF}=73\pm3\%$, where the uncertainty in NF was found by propagating the uncertainties in $A_n$ and $A_e$ in quadrature.

We checked this value of NF using the earlier method of \citet{Ritchings1976}, who generated histograms of fluence from the on-pulse region and a subset of the off-pulse region, and computed the scaling factor required to make the two comparable; they identify this scaling factor with the nulling fraction. This results in a ${\rm NF}=80\pm7\%$, consistent with the value found above.  Further observations of this pulsar to obtain high-quality pulse sequences to better characterize the single-pulse behaviour are strongly encouraged.

\begin{table}
\caption{\label{tab:timing}Observed and derived parameters 
for PSR~J0039+5407. The distance parameters were derived using the NE2025 model \citep{2026arXiv260211838O}.}

\begin{tabular}{ll}
\hline
\hline
\noalign{\smallskip}
Parameter & Value \\
\noalign{\smallskip}
\hline
\noalign{\smallskip}
Right ascension, $\alpha$ (h:m:s) (J2000) & 00:39:11.65(2) \\
Declination, $\delta$ (deg:m:s) (J2000) & 54:07:11.6(1) \\
Spin frequency, $\nu$ (Hz) &  0.446192165098(9) \\
Spin frequency derivative $\dot{\nu}$ (s$^{-2}$) & --3.12(1)$ \times 10^{-15}$ \\
Epoch of period (MJD) &  60102 \\
Dispersion measure, DM (cm$^{-3}$ pc) & 72.9(1) \\
Galactic longitude, $l$ (deg) & 121.12 \\
Galactic latitude, $b$ (deg)  & --8.71\\
50\% pulse width, $w_{50}$ (ms) & 64\\
10\% pulse width, $w_{10}$ (ms) & 92\\
\hline
DM-derived distance, $d$ (kpc) & 3.0 \\
Height above the Galactic plane, $z$ (pc) & 450\\
Characteristic age, $\tau$ (Myr) & 2.3\\
Magnetic field strength, $B$ ($10^{8}$~T) & 6.0\\
Spin-down luminosity, $\dot{E}$ ($10^{24}$~W) & 5.5\\
\noalign{\smallskip}
\hline
\end{tabular}
\end{table}

\subsection{GBP~220718}

\subsubsection{Search for low earth orbiting spacecraft}

As the number of low-Earth orbit (LEO) satellites continues to increase \citep{McDowell_2020}, distinguishing between transient RFI and genuine astrophysical signals has become vital. Indeed, a very narrow (ns duration) bright radio pulse from a decomissioned LEO satellite was
recently observed in the 700--1000~MHz band with ASKAP \citep{2025ApJ...987L..16J}. We conducted a search for LEO satellites above the Green Bank horizon at the time of GBP~220718. Using historical two-line element data from the Space-Track repository \citep{SpaceTrack}, we reconstructed the distribution of LEOs and found none within $5^{\circ}$ of the GBT boresight (Azimuth: $45.3^{\circ}$, Elevation: $77.5^{\circ}$). The closest LEO at the time of the event was COSMOS~1320, launched in 1981, at an angular separation of $7.2^{\circ}$, while the nearest Starlink satellite (STARLINK-3843, a version 1.5 spacecraft) was over $17^{\circ}$ away. While Starlink V1.5 satellites are designed for Ku-band communication, they emit unintended electromagnetic radiation in the 110--188~MHz range \citep{DiVruno_2023, 2025A&A...699A.307G}. Whether this leakage extends into L-band remains uncharacterized. However, given that these satellites would be in the far field of the GBT at these frequencies (where the beam is well-defined and sidelobe response is significantly attenuated), and given the large angular separations observed here, we consider it unlikely that LEO satellites produced GBP~220718.

\subsubsection{Search for repetitions}

The narrow-band nature of our observation of GBP~220718, and the complexity of its pulses discussed further below, motivated a series of follow-up observations to confirm it as a source of astrophysical nature, determine its origin and whether it is a repeating source. 

We followed up our initial observation of GBP~220718 with the Lovell Telescope at the Jodrell Bank Observatory (JBO). We observed the position determined in Section \ref{sec:gbpdisc} for a total of 8~h and found no significant bursts. These observations were taken on nine days over the span of four months, from October 8, 2023 to February 15, 2024. The bandwidth of the observation is 336~MHz ranging from 1396 to 1732~MHz. The RFI environment around JBO during our observations was not very clean. The majority of bursts we detected with our {\tt heimdall} pipeline were narrow-band RFI and easily recognized. Given the width of the original detection (17~ms), a system temperature of 30 K, a gain of 1~K/Jy, and a S/N of 10, we should be able to detect bursts of at least 83~mJy. Our non-detection precludes any bursts of this flux or greater during our observations.

The field of view containing GBP~220718 was observed by CHIME/FRB for a total of about 150~hours with a median 95\% completeness threshold of 3.5\,Jy\,ms \citep{2026arXiv260109399F}. No pulses were detected from this source above this fluence limit.

\subsubsection{Dispersion measure of the main pulse}

{As shown in Fig.~\ref{fig:knownFRBs}, the narrow-band nature of GBP~220718 leads to a large bow-tie feature in the DM-time plane that is consistent with a pulse with a DM of 145.5~cm$^{-3}$~pc, substantially higher than the Galactic contribution predicted by NE2025 \citep{2026arXiv260211838O} or YWM16 \citep{2017ApJ...835...29Y} which predict DMs of 55.2 and 71.9~cm$^{-3}$~pc, respectively. It is possible that GBP~220718 is actually within the Milky Way, and that the apparent DM excess is due to inaccuracies in Galactic DM models along the line of sight. We searched the ATNF catalog for any pulsars with DMs exceeding the NE2025 maximum line-of-sight DM by 90.3~cm$^{-3}$~pc or the YMW16 maximum line-of-sight DM by 73.6~cm$^{-3}$~pc, the excesses seen from this FRB. For each model,  $<1$\% of pulsars showed such large excesses, making this an unlikely explanation. In fact, most of that group of pulsars lie in globular clusters or the Magellanic Clouds, which explain many of the discrepancies.} Additionally, pulsars near GBP~220718 in the sky do not appear to have anomalously high dispersion measures, and the DM of GBP~220718 is at least four times that of every known pulsar within 10 degrees of its sky position. 

 If GBP 220718 is astrophysical, we note that it is spatially coincident with the galaxy LEDA~2277498 \citep{2003A&A...412...45P}. This candidate host, also identified as SDSS J072513.14+463143.5, lies at an angular separation of approximately $17''$ from our best-fit position, well within the estimated $1\sigma$ uncertainty. The galaxy is a faint dwarf system with a photometric redshift of $z_{\text{phot}} \approx 0.082$ \citep{2020ApJS..249....3A}. The extragalactic dispersion measure ($\text{DM}_{\text{ext}}$) of the burst further supports this association. While a crude $\text{DM}_{\text{ext}} \sim 1000z$ scaling \citep{2019A&ARv..27....4P} suggests $z \sim 0.07$, a more rigorous accounting of the foreground is required. In the direction of GBP 220718 ($l=176.3^{\circ}, b=23.4^{\circ}$), the NE2025 model predicts a Galactic disk contribution of $\text{DM}_{\text{ISM}} = 38$ pc cm$^{-3}$. Accounting for a halo contribution of $\sim 32$ pc cm$^{-3}$ \citep{Yamasaki+Totani2020} and an expected IGM contribution of $\text{DM}_{\text{IGM}} \approx 66$ pc cm$^{-3}$ at $z=0.082$ \citep{2020Natur.581..391M}, the observed DM is broadly consistent with an origin in LEDA 2277498. While these model-dependent subtractions suggest a low host contribution ($\text{DM}_{\text{host}} \lesssim 30$ pc cm$^{-3}$), the large uncertainties in the halo and IGM components preclude a precise measurement. Nevertheless, the alignment in DM-space supports the possibility that the progenitor is located in a low-density environment, perhaps in the outskirts of the galaxy.

\subsubsection{Scintillation analysis of the main pulse}

If GBP~220718 is astrophysical, it might display other effects of propagation through the interstellar and intergalactic mediums. Chief among these is diffractive scintillation, arising from inhomogeneities in the interstellar medium. Diffractive scintillation can be quantified by the diffractive timescale $\Delta t_d$ and diffractive bandwidth $\Delta\nu_d$. These quantities can be estimated by computing the two-dimensional autocorrelation function (ACF) of the pulse's dynamic spectrum and averaging over frequency and time, respectively, then fitting each one-dimensional ACF to a Gaussian or Lorentzian function and measuring the widths of the fitted functions \citep{Cordes+1985,Cordes1986}. In particular, $\nu_d$ is calculated as the half width at half maximum of the fitted Gaussian or Lorentzian.

Efforts have been made to estimate $\Delta \nu_d$ from observable quantities. \citet{Bhat+2004} fit a relation between the scattering timescale $\tau_s$, dispersion measure $\mathrm{DM}$ and observing frequency $f$:
\begin{equation}
\log\tau_s=-6.46+0.154\log(\mathrm{DM})+1.05(\log\mathrm{DM})^2-3.86\log f,
\end{equation}
where $\tau_s$ is in milliseconds, $\mathrm{DM}$ is in ~cm$^{-3}$~pc and $f$ is in GHz. Assuming a DM of 145.5~cm$^{-3}$~pc and a central observing frequency of 1.110 GHz, we find an expected scattering timescale for GBP~220718 of approximately 0.05 ms. For a Kolmogorov spectrum, 
\begin{equation}
\Delta\nu_d=\frac{1.16}{2\pi\tau_s}
\end{equation}
yielding an expected diffractive bandwidth of approximately 0.366 kHz. Unfortunately, this is significantly lower than the GREENBURST frequency channel bandwidth of 234~kHz, indicating that measuring $\Delta \nu_d$ for GBP~220718 may not be possible.

\begin{figure}
\centering
    \includegraphics[width=0.45\textwidth]{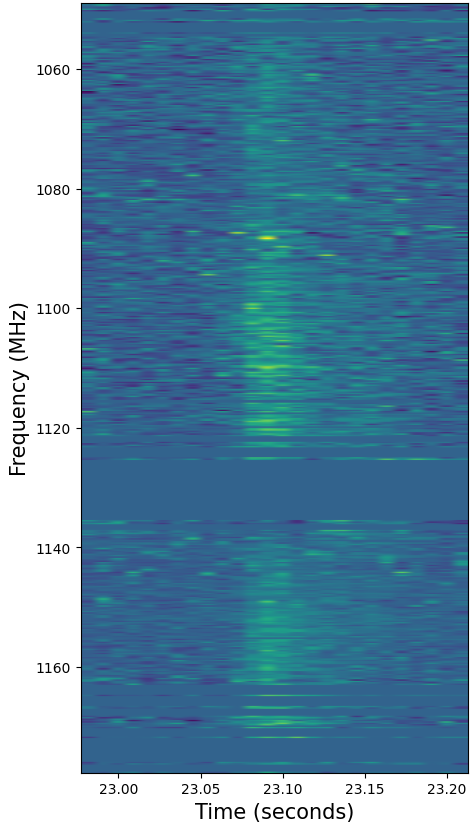}
    \caption{Dynamic spectrum of the main pulse of GBP~220718. Channels on the outside of the band were excised and  not used to compute the ACF.}
    \label{fig:Burst_2}
\end{figure}

Given that there is a significant amount of scatter in the observed DM-$\tau_s$ relation, we still attempt to measure a diffractive bandwidth for GBP~220718. We isolated the three largest features in the dynamic spectrum of the FRB, computed the one-dimensional frequency ACF for each subburst by averaging the two-dimensional ACF in time, and fit Lorentzians to each ACF, as the ACFs display strong wings. We arrived at measurements of $\Delta\nu_d$ of $11\pm4$~MHz, $7.4\pm2.5$~MHz and $7.9\pm2.8$~MHz, respectively. As an example, Fig.~\ref{fig:Burst_2} shows the dynamic spectrum of the second of the three components used.

All three fits have difficulty fitting the central portion of the ACF, and so these widths are likely overestimates of the true diffractive bandwidth. Even accounting for this, the results are significantly higher than expected by the \citet{Bhat+2004} relation. A closer examination of the dynamic spectra of the subbursts suggests that the fits may be absorbing spectral features of the burst, rather than actual diffractive scintillation.

\subsubsection{Associated emission around the main pulse}

\begin{figure*}
    \centering
    \includegraphics[width=0.9\textwidth]{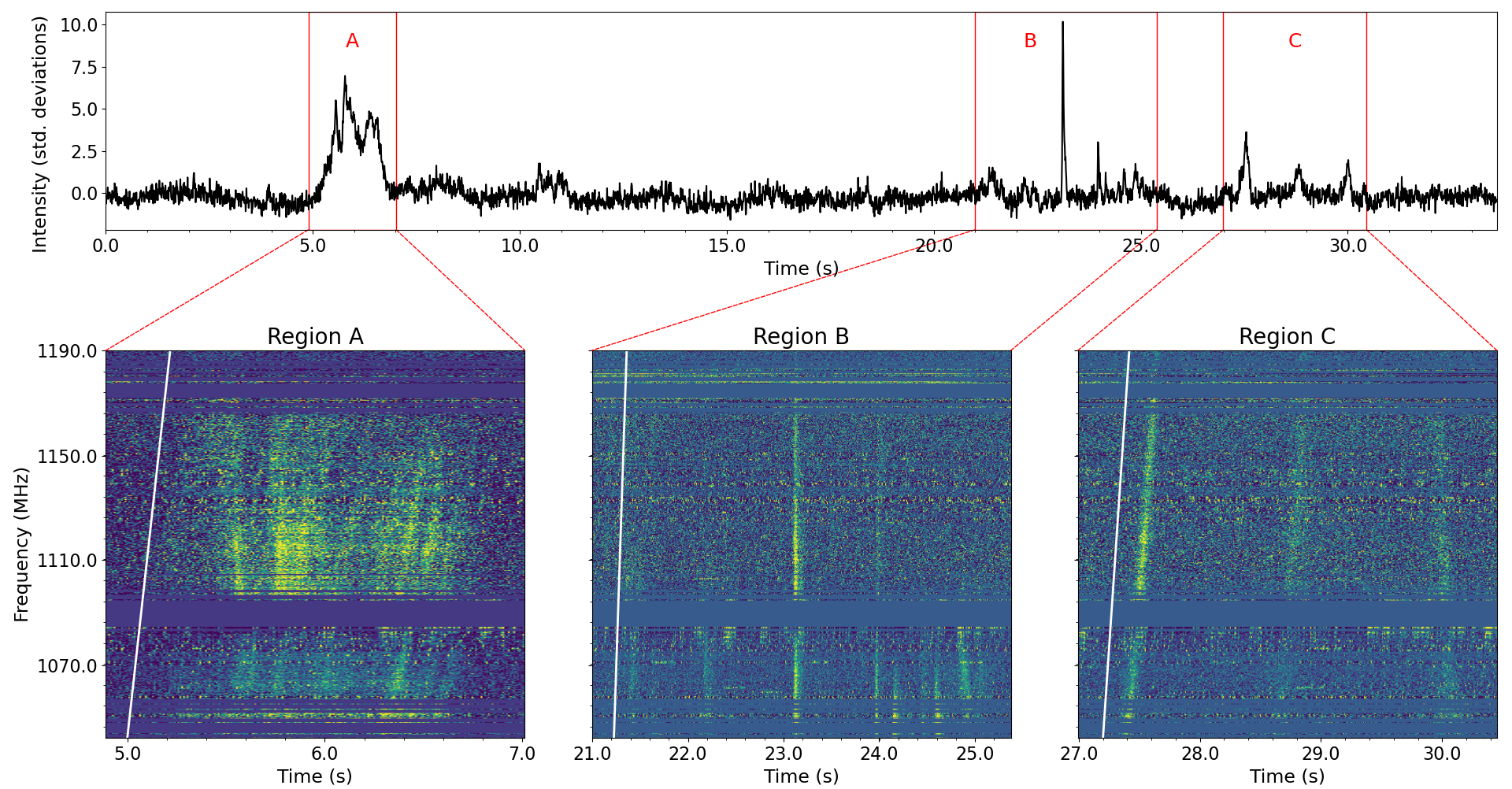}
    \caption{Summary of the observation containing all the emission recorded from GBP~220718, cleaned with the composite filter described in \citep{millisecond_filters} with weighted zero-DM subtraction. The top panel shows the dedispersed time series referenced to a DM of 145.5~cm$^{-3}$~pc. Three different regions are shown, referred to as A, B and C. The bottom panels show waterfall plots also dedispersed to a DM of 145.5~cm$^{-3}$~pc zooming in around each region. The original detection shown in Fig.~\ref{fig:knownFRBs} is the bright pulse at the centre of region B. The white lines show the expected behaviour for a zero-DM signal.}
    \label{fig:gallery}
\end{figure*}

Each time a significant pulse like GBP~220718 is recorded, a filterbank file corresponding to the entire block of data (9 minutes) around that event is archived for offline analysis. Our examination of the data around GBP~220718 does reveal other bursts of emission in this same band, as shown in Fig.~\ref{fig:gallery}. In addition to the bright central pulse, Fig.~\ref{fig:gallery} shows several less luminous components that were present during the time the source was in the field of view of the telescope ($\approx$~40~s), but not initially detected by our real-time pipeline. As Fig.~\ref{fig:gallery} shows, pulses are detected in three main regions of the time series which we label A, B, and C. Also shown are waterfall plots which show the frequency structure in these three regions. In this display, only the 100~MHz band in which GBP~220718 was detected (Fig.~\ref{fig:knownFRBs}) is shown and all data are de-dispersed to a DM of 145.5~cm$^{-3}$~pc. Some of these components have similar spacing in time, especially in region B (where the main pulse was initially detected) and the spacing is approximately 0.85~s. A periodicity analysis on the data collected for the time series shown in Fig.~\ref{fig:gallery}, as well as region B alone using PRESTO \citep{Ransom-2011} and \textsc{riptide} \citep{Morello_2020} did not yield any significant periodicities. 

As Fig.~\ref{fig:gallery} shows, while there clearly are other pulses present in the observation at the same DM of GBP~220718, there are a number of features in the same narrow frequency band that are consistent with DMs of zero. Such features are most prominent in region A which also shows a number of vertical bands (i.e., DM values around 145.5~cm$^{-3}$~pc) alongside sloping features that correspond to sub-pulses consistent with zero DM. While a variation of slopes in multi-component FRBs have been seen in CHIME/FRB data and described in detail in \citet{2024ApJ...974..274F}, the features we observe in Fig.~\ref{fig:gallery} show a much greater variation in DM over a much narrower band than seen by CHIME/FRB. If the emission shown here is truly astrophysical in nature, it would be unlike any other pulses seen to date. 

\begin{figure*}
    \centering
     \includegraphics[width=0.49\textwidth]{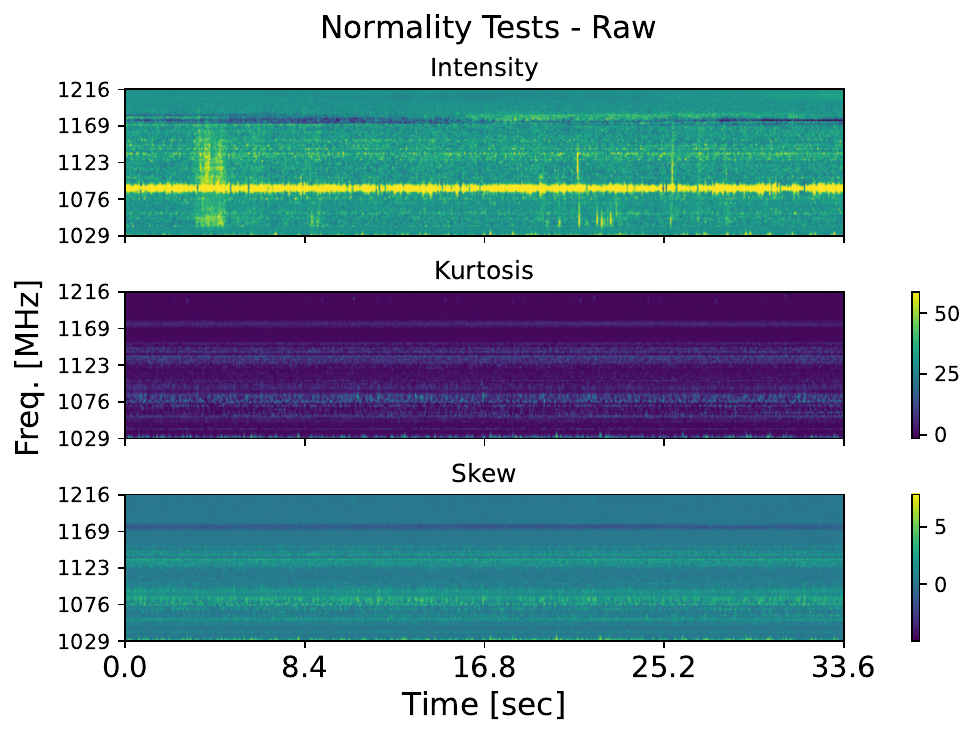}
    \includegraphics[width=0.49\textwidth]{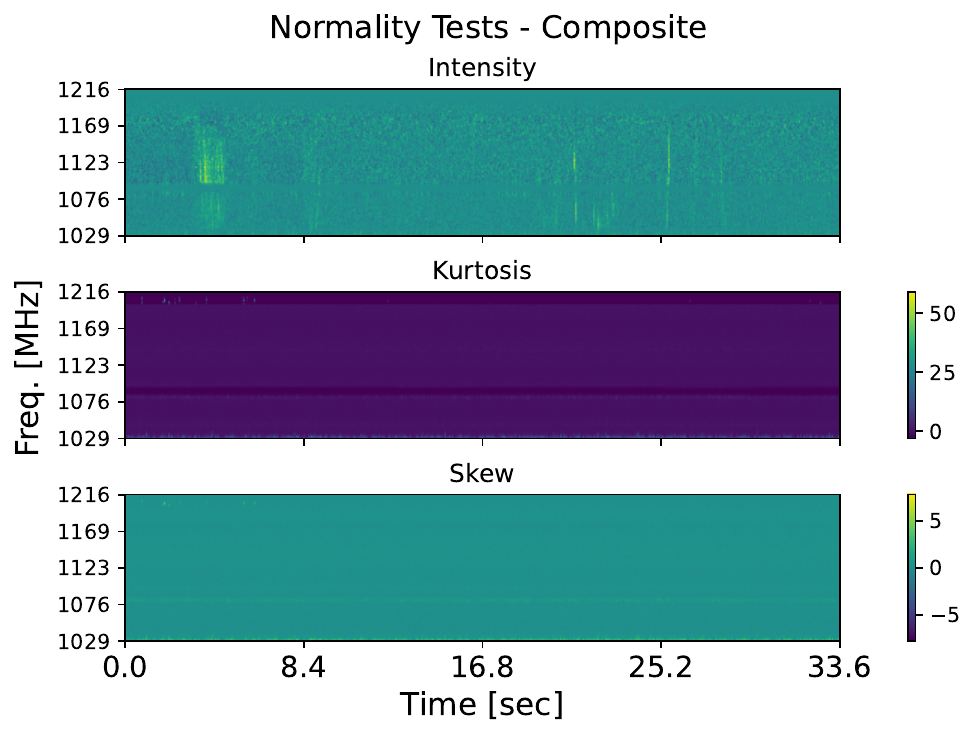}
    \caption{Kurtosis and skew analyses on blocks of 64 times samples for the data containing GBP~220718. The left hand plots show the data without any RFI filtering, the right hand plots are post cleaning with the composite filter \citep{millisecond_filters} with weighted zero-DM subtraction. For both cases, we show the dynamic spectra (top), kurtosis values (middle) and skew values (bottom). As the right-hand panels  show, there is no significant kurtosis or skew for GBP~220718.}
    \label{fig:normality}
\end{figure*}

While the initial pulse seen for GBP~220718 shows a number of FRB-like properties, the presence of the additional pulses shown in Fig.~\ref{fig:gallery}, many of which exhibit no dispersion, while still falling in the same narrow frequency range, leads us to be skeptical of an astrophysical origin for this source. On the other hand, the fact that it passed our RFI filtering described in \S \ref{sec:obsover}, which flags signals that are highly non-Gaussian means that we cannot safely conclude 
a terrestrial origin for the signal. The kurtosis and skew values for raw and RFI-filtered dynamic spectra are shown in Fig.~\ref{fig:normality}. Emission from GBP~220718 is visible in both intensity data, but without any corresponding increase in either kurtosis or skew. This is in contrast to the RFI bands visible in the left hand side, which show brightly in kurtosis and skew. In the absence of further information that would possible from an array receiver, we cannot currently rule out from this single-beam observation of GBP~220718 with GREENBURST whether signal is terrestrial in nature. Our experiences with GBP~220718 are consistent with the difficulties in classifying FRBs as discussed by \citet{2018MNRAS.481.2612F}. 

\section{Conclusions}

The first 20,000 hours of GREENBURST observations resulted in the detections of 50 pulsars, one of which (PSR~J0039+5407) was previously unknown, three previously known FRBs and one dispersed pulse of unknown origin (GBP~220718). PSR~J0039+5407 is a 2.2~s pulsar with a nulling fraction on the order of 80\%. GBP~220718 was originally detected as a narrow-band signal with a DM of 145.5~cm$^{-3}$~pc. Based on FRB observations to date, such properties are consistent with a nearby repeating FRB. However, in spite of significant follow-up observations with other telescopes, no additional events were found. In addition, upon closer inspection of the data taken around the epoch of detection for GBP~220718, however, we found a number of additional pulses that are consistent with zero DM signals in the same part of the band. 
 We searched for possible LEO satellites above the horizon at the time of the observation of GBP~220718, but found none. In addition,
the higher order moments of GBP~220718 are inconsistent with most known sources of terrestrial origin.  If astrophysical, GBP~220718 appears to be coincident with the nearby galaxy LEDA 2277498. Further observations are necessary to make a firm conclusion as to whether GBP~220718 is astrophysical or terrestrial in nature.

GREENBURST observations currently underway include a single-pulse study of all the millisecond pulsars we have observed to date (Tabassum et al., in preparation) and a survey of nearby elliptical galaxies for FRBs (Paine et al., in preparation). Ongoing work is focused on extending the range of pulse widths searched to improve the sensitivity of GREENBURST to long-period transients \citep[see, e.g.,][]{2023Natur.619..487H}, providing 
sensitivity to events such as the one seen in region A of Fig.~\ref{fig:gallery}.

\section*{Acknowledgements}

The Green Bank Observatory is a facility of the National Science Foundation operated under cooperative agreement by Associated Universities, Inc. We thank Green Bank Observatory staff Ryan Lynch and Evan Smith for their assistance with the followup observations. GREENBURST operations have been supported by NSF Astrononomy and Astrophysics awards 1616042 and 2406570. S.P. and D.R.L. acknowledge support from NSF award 2406570. S.T. and D.R.L. acknowledge support from NSF award 2307581. S. S.  and K. E. H. are WVU Ruby Doctoral Fellows. M. F. acknowledges support from the WV Spacegrant Consortium. E. J. M. acknowledges NSF support from the WVU Research Experiences for Undergraduates program under award 2348764. F.A.D is supported by an NRAO Jansky Fellowship. We acknowledge that CHIME is located on the traditional, ancestral, and unceded territory of the Syilx/Okanagan people. We are grateful to the staff of the Dominion Radio Astrophysical Observatory, which is operated by the National Research Council of Canada. CHIME operations are funded by a grant from the NSERC Alliance Program and by support from McGill University, University of British Columbia, and University of Toronto. CHIME was funded by a grant from the Canada Foundation for Innovation (CFI) 2012 Leading Edge Fund (Project 31170) and by contributions from the provinces of British Columbia, Québec and Ontario. The CHIME/FRB Project, which enabled development in common with the CHIME/Pulsar instrument, was funded by a grant from the CFI 2015 Innovation Fund (Project 33213) and by contributions from the provinces of British Columbia and Québec, and by the Dunlap Institute for Astronomy and Astrophysics at the University of Toronto. Additional support was provided by the Canadian Institute for Advanced Research (CIFAR), the Trottier Space Institute at McGill University, and the University of British Columbia. The CHIME/Pulsar instrument hardware is funded by the Natural Sciences and Engineering Research Council (NSERC) Research Tools and Instruments (RTI-1) grant EQPEQ 458893-2014. We thank Matthew Stevenson for assistance with the LEO satellite analysis and the referee for very useful comments on an earlier version of the manuscript.

\section*{Data Availability}

All GREENBURST triggers are available from the authors upon reasonable request. Processed data from the pipeline for detections of known pulsars
are available via the GREENBURST website: https://greenburstwvu.github.io.


\bibliographystyle{mnras}
\bibliography{refs} 


\bsp	
\label{lastpage}
\end{document}